\def\year{2022}\relax
\documentclass[letterpaper]{article} 
\usepackage{aaai22}  
\usepackage{times}  
\usepackage{helvet}  
\usepackage{courier}  
\usepackage[hyphens]{url}  
\usepackage{graphicx} 
\usepackage{natbib}  
\usepackage{caption} 
\DeclareCaptionStyle{ruled}{labelfont=normalfont,labelsep=colon,strut=off} 
\usepackage{amsmath}
\usepackage[ruled,linesnumbered,noresetcount]{algorithm2e}
\usepackage{atbegshi}
\AtBeginDocument{\AtBeginShipoutNext{\AtBeginShipoutDiscard}}
\usepackage{newfloat}
\usepackage{listings}
\title{U-shaped Transformer with Frequency-Band Aware Attention for Speech Enhancement}
\author{
    Yi Li, \textsuperscript{\rm 1}
    Yang Sun, \textsuperscript{\rm 2}
    Syed Mohsen Naqvi \textsuperscript{\rm 1}
}
\affiliations{
    \textsuperscript{\rm 1} Newcastle University \textsuperscript{\rm 2} University of Oxford\\
    y.li140, mohsen.naqvi@newcastle.ac.uk, Yang.sun@bdi.ox.ac.uk\\
    
}
\usepackage{bibentry}

\begin{document}
\mbox{}
\thispagestyle{empty}
\newpage
\maketitle
\begin{abstract}
The state-of-the-art speech enhancement has limited performance in speech estimation accuracy.  Recently, in deep learning, the Transformer shows the potential to exploit the long-range dependency in speech by self-attention. Therefore, it is introduced in speech enhancement to improve the speech estimation accuracy from a noise mixture. However, to address the computational cost issue in Transformer with self-attention, the axial attention is the option i.e., to split a 2D attention into two 1D attentions. Inspired by the axial attention, in the proposed method we calculate the attention map along both time- and frequency-axis to generate time and frequency sub-attention maps. Moreover, different from the axial attention, the proposed method provides two parallel multi-head attentions for time- and frequency-axis. Furthermore, it is proven in the literature that the lower frequency-band in speech, generally, contains more desired information than the higher frequency-band, in a noise mixture. Therefore, the frequency-band aware attention is proposed i.e., high frequency-band attention (HFA), and low frequency-band attention (LFA).  The U-shaped Transformer is also first time introduced in the proposed method to further improve the speech estimation accuracy. The extensive evaluations over four public datasets, confirm the efficacy of the proposed method.

\end{abstract}

\section{1\quad Introduction}
Speech enhancement, aiming to improve the desired speech quality, is a crucial topic of audio signal processing and has been used in many real-world applications, including automatic speech recognition (ASR), teleconferencing, and robotics \cite{Luo}. Because the importance of the research aspect, numerous deep learning approaches have been proposed to solve the problem and improve performance \cite{aaai, ae, yiluo}. However, existing deep learning techniques and models typically cause different levels of degradation on the enhancement performance \cite{NTT, intro, CSA1}.


Some novel network models are proposed and extensively used in speech enhancement, which show high competitiveness compared with conventional signal processing techniques. As the outstanding performance in natural language processing (NLP), the Transformer is introduced in speech enhancement problem \cite{setf}. The Transformer follows the encoder and decoder architecture with stacked self-attention and point-wise feed-forward layers \cite{DT}. However, the Transformer has a limitation as difficult to process a multiple-sequence alignment as an input for feature extraction in the inference stage \cite{survey}. Thus, the novel Transformer model interleaves attention map in weights and heights of an alignment as in axial attention, enabling it to extract features during the inference phase to reduce computation complexity \cite{ATMT}. In our work, the attention maps are constructed as time and frequency directions. Then, skip connections are introduced in Transformer to reduce the loss of feature information at each convolution \cite{bunet}. Besides, in the sub-layers of the proposed U-shaped Transformer, simplified as U-Transformer, multi-head self-attention blocks are exploited for parallel and independent computations.

\begin{figure*}[htbp!]
\centering
\includegraphics[width=16cm, height=9cm]{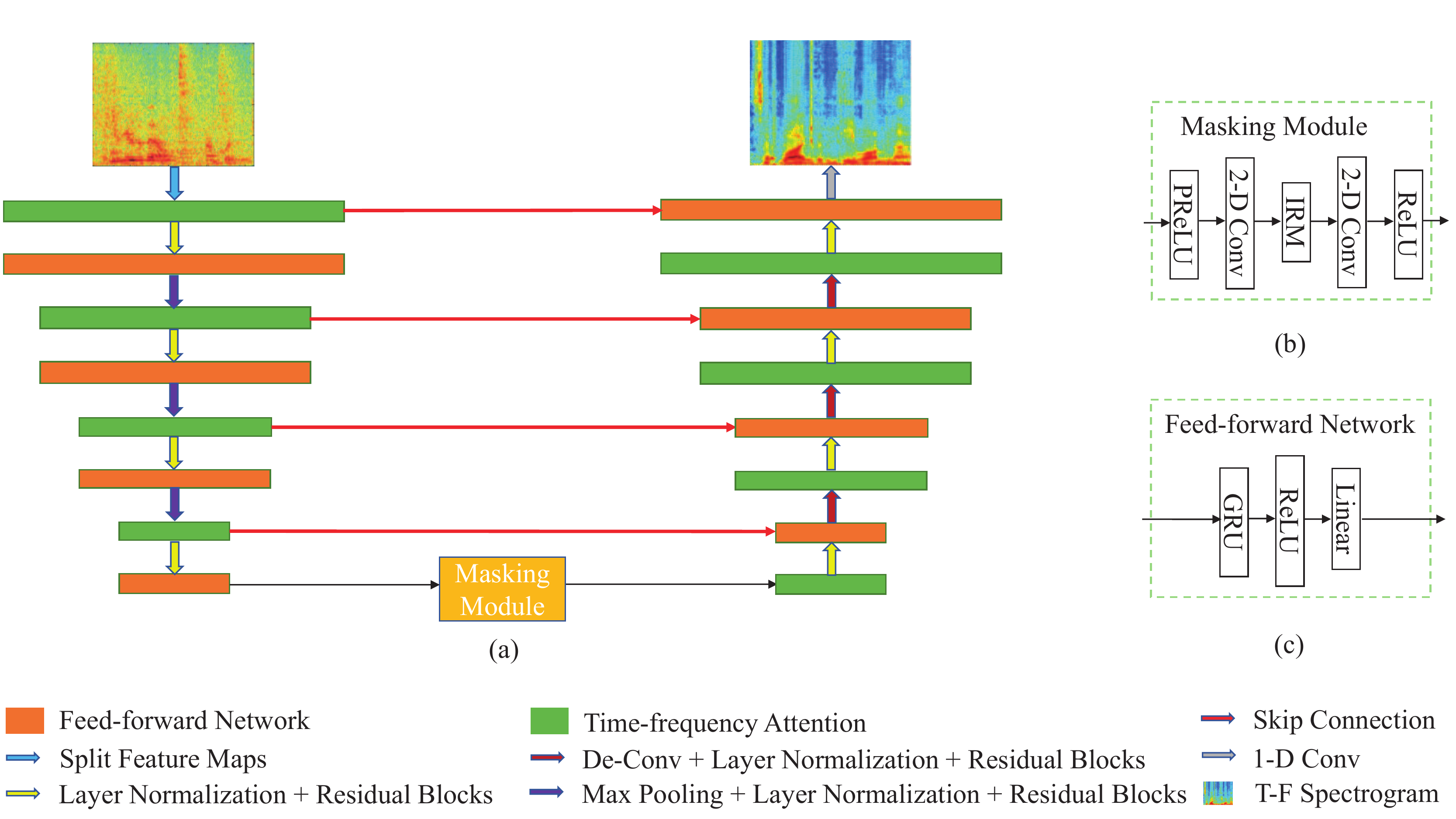}
\caption{The overall architecture of the proposed time-frequency attention based U-Transformer. As the input of the encoder, extracted features from the noisy spectrogram are split into time- and frequency-axis, and masked by multi-head attentions. After the feature is recovered by the de-conv layers in the decoder, the reconstructed spectrogram is obtained as the output. The masking module and feed-forward network are presented in (b) and (c), respectively.}\centering
\end{figure*}
Most of deep learning architectures for speech enhancement are formulated in the full-band time-frequency (T-F) representation of the speech mixture \cite{tf1, tf2, tf3}. By using short-time Fourier transform (STFT), the state-of-the-art methods estimate the spectrogram of the desired speech signal from the mixture spectrogram \cite{stft} \cite{stft1}. However, it has been confirmed that the background noise is uniformly distributed at the full band and human speech occupies in the lower frequency-band \cite{IET}. Thus, the whole T-F attention map is further divided into three sub attention maps, time attention (TA), high frequency-band attention (HFA), and low frequency-band attention (LFA). Because the lower band occupies the 0-4000 Hz consists of almost all the speech signals in the mixture, the LFA is paid more weights such as 16-head attention and different learnable vectors for keys, values, and queries to fully use the desired source information, while the HFA is only trained with light weights and an overall learnable vector to improve parameter efficiency.

The contributions of this paper are summarized as follows:

$\bullet $  A U-shaped transformer is first time introduced to address the speech enhancement problem. The skip connections are also added between the sub-layers in the encoder and decoders.

$\bullet $   The 2-D attention map is split into two 1-D time and frequency sub-attention maps, which allow the parallel calculations to facilitate the training. Independent learnable vectors for query, keys and values are exploited for the information between the positions at the sub-attention maps.

$\bullet $  The multi-head attentions in time- and frequency-axis are further refined into three attentions i.e., time attention (TA), high frequency-band attention (HFA), and low frequency-band attention (LFA). Furthermore, different number of heads and learnable vectors are also used for the three multi-head attentions to efficiently train the sub-attention maps.
\section{2\quad Related Work}

Different from recurrent neural networks (RNNs), the Transformer processes the input in parallel and does not necessarily depend on the previous input to be processed. As aforementioned, the Transformer adopts the scaled dot-product attention and treats the encoded representation of the input as a set of keys $K$ and values $V$ \cite{atyn}. In the decoder, the previous output is compressed into a query $Q$ and the next output is constructed by mapping the query and the set of keys and values. Thus, the output is a weighted summation of the values, where the assigned weight of each value is determined by the dot-product of the query with all the keys \cite{atyn}.

Rather than only estimating the attention block once, the scaled dot-product attention is utilized for the parallel calculation of the multi-head attention multiple times \cite{tfeer}. The independent attention outputs are simply concatenated and linearly transformed into the expected dimensions \cite{st1}. In addition, for each multi-head attention block, the feed-forward layer is comprised followed by a layer normalization to process the output from the attention layer to rescale and better fit the input for the next sub-layer \cite{st2}. Although two linear transformations exploited in the feed-forward layer share the same architecture across different positions, they employ different parameters between two layer normalizations \cite{st3}. The overall architecture of the proposed U-Transformer is presented in Figure 1 and color version of the figure is better to understand.

The main idea is supported by the basic theory in speech processing that compared to noise interference, mostly of the human speech energy exists in the lower frequency-band \cite{voice}. In telephony, the averaged frequency range for human speech varies from approximately 500 Hz to 2000 Hz. The allocated bandwidth for a monaural voice-frequency transmission is around 4000 Hz. Figure 2 is the spectra of mixtures with various noise interferences randomly selected from the NOISEX dataset \cite{noise}.

\begin{figure}[h!]
\centering
\includegraphics[width=8.5cm, height=5cm]{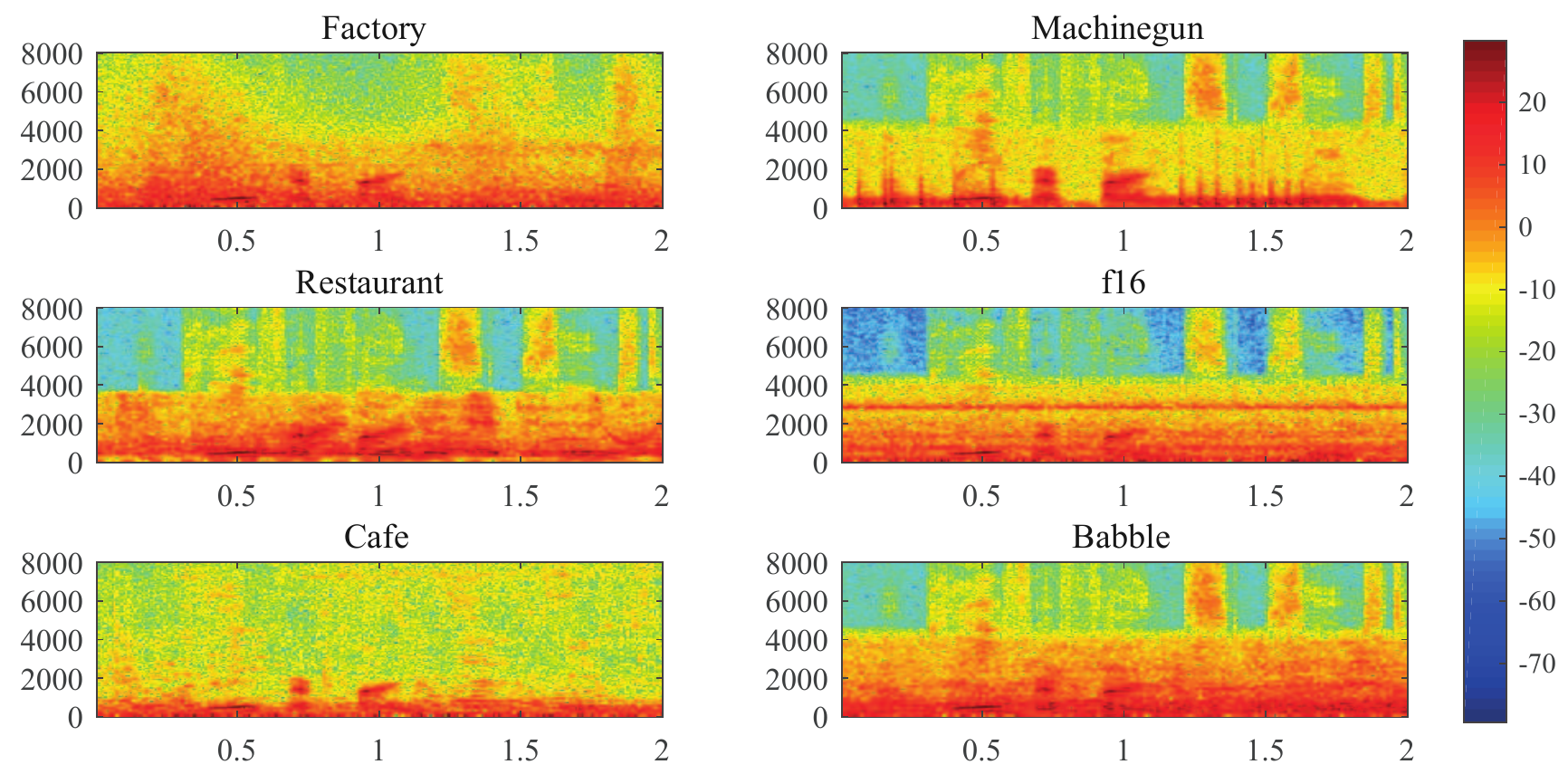}
\caption{The spectra of mixtures with various noise interferences from the NOISEX database at -5 dB SNR level. The x- and y-axis are time (s) and frequency (Hz), respectively.}\centering
\end{figure}

From Figure 2, it can be observed that the mixture energy, intensively, distributes at the lower frequency-band [0-4000 Hz]. Therefore, the LFA is paid with more computation costs in the proposed frequency-band aware attention to further improve performance.

\section{3\quad Proposed Methods}
In this section, we present the T-F attention based U-Transformer with the frequency-band aware method. Each block in the overall architecture of the network is introduced in the first subsection. Followed by the description of encoder-decoder U-Transformer, T-F attention and frequency-band aware attention with different multi-head attention parameters are presented.
\subsection{Speech Enhancement U-Transformer}
The overall model architecture of the proposed T-F attention-based U-shaped Transformer is presented in Figure 1. We assume the input $\mathcal{X}^{t}=\left\{\mathbf{x}_{1}^{t}, \ldots, \mathbf{x}_{L}^{t} \right\}$ at time $t$, and the aim of speech enhancement problem is to estimate the desired speech signal $\mathcal{Y}^{t}=\left\{\mathbf{y}_{1}^{t}, \ldots , \mathbf{y}_{L}^{t}\right\}$ for $L$-length inputs/outputs where $\mathbf{x}_{L}^{t}$ represents the $L$-$th$ frame of the magnitude spectrogram of noisy speech.

Initially, the noisy speech is generated with the clean speech signal and the noise interference. The segmentation layer splits $\mathcal{X}^{t}$ into $L$ frames. The encoder consists of four Transformer blocks and each block has a T-F attention, a feed-forward network, and two layer normalizations. 

The conventional Transformer encoder consists of three important modules: positional encoding, multi-head attention, and position-wise feed-forward network. In this work, the encoder is comprised of four sub-layers and each encoder layer has a multi-head attention and a feed-forward network. The residual connection \cite{rc} is exploited in both multi-head attention mechanism and feed-forward network. In the proposed T-F attention method, two multi-head attention blocks are applied on sub attention maps to extract the desired information at time- and frequency-axis, respectively. The feed-forward network estimates the compatibility function with a single hidden layer \cite{atyn}. Different from the conventional Transformer encoder, in the feed-forward network, the first fully connected layer is replaced by a gated recurrent unit (GRU) \cite{gru} layer because the GRU has a simpler structure and trains faster than local LSTMs. Moreover, the same dimension of attentions maps is obtained from the input and output of one sub-layer in the U-Transformer, e.g., $d_{\text {layer }}=512$ in the first sub-layer, to facilitate the residual connections. A layer normalization is followed by both the multi-head attention and the feed-forward network on the addition of before and after the operations.

The output from the encoder stack provides to the masking module which consists of two 2-D convolutional layers with ReLU and PReLU activation functions. Because information is propagated along time- and frequency-axis, to utilize the conditioning information, i.e., the relationship between each time and frequency point, the masking module is exploited between the encoder and decoder. The encoded representation is shifted up for the causality of the conditioning information with the ideal ratio mask (IRM) to estimate the target speech signal from the noisy mixture as \cite{IRM1}:
\begin{equation}
I R M=\left(\frac{S^{2}}{S^{2}+N^{2}}\right)^{\beta}
\end{equation}
where $S^{2}$ and $N^{2}$ are speech energy and noise energy, respectively, and the tunable parameter $\beta$ is set to 0.5. In the sub-layer of the decoder, the feed-forward network obtains the masked attention maps from two inputs: (1) The attention map from the corresponding sub-layer in the encoder is introduced by a skip connection. The skip connection plays an important role in reducing the loss of feature information at each convolution \cite{bunet}. (2) The output from the T-F attention block of the decoder. A concatenation operation is applied in the feed-forward network at each sub-layer of the decoder to integrate two inputs. Compared with the conventional Transformer \cite{atyn}, the proposed U-Transformer benefits from the concatenation as combining the particular information from different compression levels to achieve promising performance on speech enhancement \cite{unet2}. Moreover, similar as the encoder, a layer normalization and residual blocks are added followed by both the multi-head attention and the feed-forward network. The enhanced speech can be obtained from the output layer after the 1-D convolutional layer.

\subsection{Time-frequency Attention}
In the proposed U-Transformer, a self-attention layer is implemented in each sub-layer and takes as input an L-length sequence of embeddings and produces a same size output sequence. The output of the attention matrix is presented as \cite{atyn}:
\begin{equation}
\mathcal{A}(Q, K, V)=\operatorname{Softmax}\left(\frac{Q K^{\top}}{\sqrt{L}}\right) V
\end{equation}
where $\top$ is the symbol for transpose. Initially, the full attention map as $T\times F\times d_{\text {layer}}$ is input to each sub-layer of the encoder. Different from the axial attention \cite{ATMT}, due to the property of the speech signal, the proposed T-F attention splits the 2-D attention map into two 1-D sub maps in time- and frequency-axis as $T\times d_{\text {layer}}$ and $ F\times d_{\text {layer}}$, respectively. Each sub attention map propagates information along one specific axis. Moreover, two sub attention maps are constructed as parallel calculations on multiple GPUs to optimize the training process. The multi-head attention for the time direction can be presented as:
\begin{equation}
\begin{aligned}
\operatorname{multihead}(Q^{t}, K^{t}, V^{t}) =\left[\operatorname{h}_{1} ; \ldots ; \text { h }_{8}\right] W_{t}^{O} \\
\text { where h }_{i} =\mathcal{A}\left(Q^{t} W_{t}^{Q}, K^{t} W_{t}^{K}, V^{t} W_{t}^{V}\right)
\end{aligned}
\end{equation}
where $W$s are the weights required to be trained and $W_{t}^{O} \in R^{8\times d_{v} \times d_{\text {layer }}}$, $W_{t}^{Q},  W_{t}^{K} \in R^{d_{\text {layer }} \times d_{k}} $, $ W_{t}^{V} \in R^{d_{\text {layer}} \times d_{v}}$. In the time attention map, query, keys and values are presented as $Q^{t}, K^{t}, V^{t}$. The number of the head is presented as $i$ and set to 8 in the proposed T-F attention method similar to the original Transformer \cite{atyn}. Similarly, for frequency attention, we use same equations but different notation $f$:

\begin{equation}
\begin{aligned}
\operatorname{multihead}(Q^{f}, K^{f}, V^{f}) =\left[\operatorname{h}_{1} ; \ldots ; \text { h }_{8}\right] W_{f}^{O} \\
\text { where h }_{i} =\mathcal{A}\left(Q^{f} W_{f}^{Q}, K^{f} W_{f}^{K}, V^{f} W_{f}^{V}\right)
\end{aligned}
\end{equation}
The multi-head attention mechanism shows high efficacy to learn the long-term dependencies since a direct connection between every pair of positions is used. The weights of the multi-head attention layer are computed by pooling over the query-key affinities $Q_{p}^{t} K_{p}^{t}$, key-dependent positional bias term $K_{c}^{t} r_{c-p}^{K^{t}}$:
\begin{equation}
W_{t}^{Q} = \sum_{c \in \mathcal{N}_{1 \times n}(c)} (Q_{p}^{t} K_{p}^{t}+K_{c}^{t}r_{c-p}^{K^{t}})r_{c-p}^{V^{t}}
\end{equation}
\begin{equation}
W_{t}^{K} = \sum_{c \in \mathcal{N}_{1 \times n}(c)} Q_{p}^{t} r_{c-p}^{Q^{t}}r_{c-p}^{V^{t}}
\end{equation}
\begin{equation}
W_{t}^{V} = \sum_{c \in \mathcal{N}_{1 \times n}(c)} (Q_{p}^{t} K_{p}^{t}+Q_{p}^{t} r_{c-p}^{Q^{t}}+K_{c}^{t} r_{c-p}^{K^{t}})V_{c}
\end{equation}
where $\mathcal{N}_{1 \times n}(c)$ is the local $1 \times n$ region around position $c$. The location compatibility is estimated by the inner product $Q_{p}^{t} r_{c-p}^{Q^{t}}$ between locations $(p, c)$. Then, $r_{c-p}$ is learnable vector to update the weights and the superscripts refer to the multi-head attention parameters.

The outputs of time and frequency attentions can be written as:
\begin{equation}
M_{t}=\operatorname{multihead}(Q^{t}, K^{t}, V^{t})
\end{equation}
\begin{equation}
M_{f}=\operatorname{multihead}(Q^{f}, K^{f}, V^{f})
\end{equation}
Then, the masked attention maps are integrated and processed by a feed-forward network to obtain the output of the improved Transformer decoder at time $t$, where residual connections and layer normalization $h(\cdot  )$ are added as well.
\begin{equation}
\mathbf{y}_{i}^{t}= \operatorname{ReLU}(h(M_{t}+M_{f}+M_{p} )) W_{i}+b_{i} 
\end{equation}
where the $i$-$th$ weight $W_{i} \in R^{d_{layer} \times T}$ and $i$-$th$ bias $ b_{i} \in R^{T}$ are trained with the output from the previous layer $M_{p}$. The desired speech signal $\hat{\mathcal{Y}}^{t}$ is estimated by integrating $L$ frames. The pseudo-code of the proposed T-F attention is summarized as Algorithm 1.

\makeatletter
\newcommand{\AlgoResetCount}{\renewcommand{\@ResetCounterIfNeeded}{\setcounter{AlgoLine}{0}}}
\newcommand{\AlgoNoResetCount}{\renewcommand{\@ResetCounterIfNeeded}{}}
\SetAlgoNoLine
\LinesNumberedHidden
\makeatother
\lstset{%
	basicstyle={\footnotesize\ttfamily},
	numbers=left,numberstyle=\footnotesize,xleftmargin=2em,
	aboveskip=0pt,belowskip=0pt,%
	showstringspaces=false,tabsize=2,breaklines=true}


\begin{algorithm}

  
  \SetKwInOut{Input}{input}\SetKwInOut{Output}{output}

  \Input{Extracted feature map as $T\times F\times d_{\text {layer}}$, learning rate $\eta$, epoch $E_{\max }$ }
  \Output{Attention map as $T\times F\times d_{\text {layer}}$}
  \BlankLine
  Initialize learning vectors\;
  \For{$E = 1, 2, ...,$  $E_{\max }$}{
    \For{each row $t \in[1, T]$}{
      Train $r_{c-p}^{Q^{t}}$, $r_{c-p}^{K^{t}}$ and $r_{c-p}^{V^{t}}$ \;
      $W_{t}^{O}$, $W_{t}^{Q}$, $ W_{t}^{K}$, $W_{t}^{V}\leftarrow r_{c-p}^{Q^{t}}$, $r_{c-p}^{K^{t}}$, $r_{c-p}^{V^{t}}$ for Eq. (3)\;
      Update the time attention map $M_{t}$\;
    }
    \For{each column $f \in[1, F]$}{
      Train $r_{c-p}^{Q^{f}}$, $r_{c-p}^{K^{f}}$ and $r_{c-p}^{V^{f}}$\;
      $W_{f}^{O}$, $W_{f}^{Q}$, $ W_{f}^{K}$, $ W_{f}^{V}\leftarrow r_{c-p}^{Q^{f}}$, $r_{c-p}^{K^{f}}$, $r_{c-p}^{V^{f}}$ for Eq. (4)\;
      Update the frequency attention map $M_{f}$\;
    }
    
    $M = M_{f}+M_{t}$ \;
      $\mu_{i}\leftarrow x_{i j}$    \qquad    //mini-batch mean \;
    $\sigma_{i} \leftarrow  x_{i j}, \mu_{i}$   \qquad    //mini-batch variance\;
    $\hat{x_{i j}} \leftarrow x_{i j},\mu_{i},\sigma_{i},\epsilon$  \     //normalize with error $\epsilon$\;
    
      $M_{sum} = M + M_{p}$ \;
     
   }
  \caption{Time-frequency Attention Algorithm.}\label{algo_disjdecomp}
\end{algorithm}

\subsection{Frequency-band Aware Attention}
As aforementioned, to fully use the desired speech information, T-F attention is further divided into three multi-head attentions, time attention (TA), high frequency-band attention (HFA), and low frequency-band attention (LFA), as shown in Figure 3.
\begin{figure}[h!]
\centering
\includegraphics[width=8cm, height=8cm]{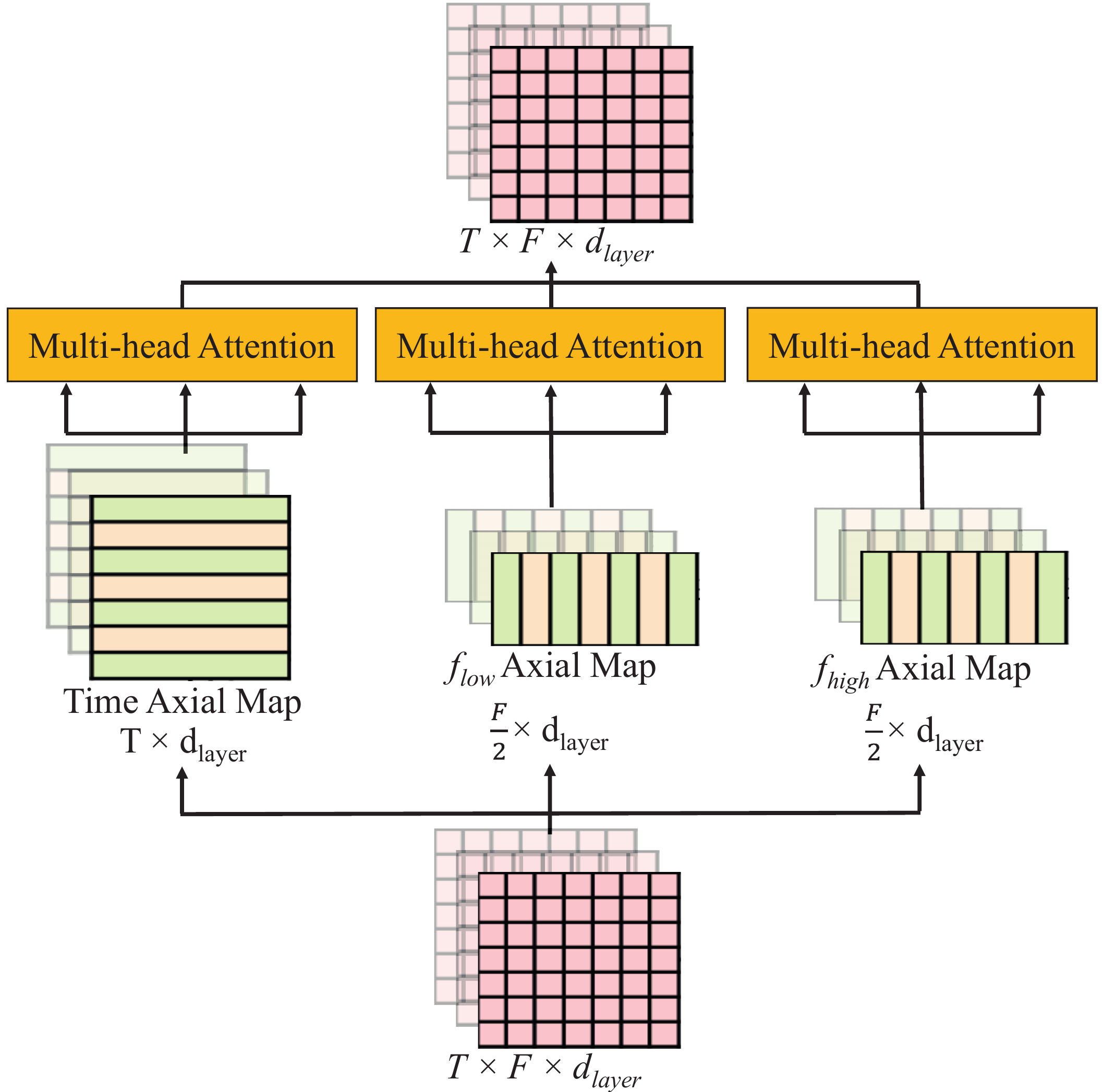}
\caption{The frequency-band aware attention. The input and output maps have the same size as $T\times F\times d_{\text {layer}}$ at each sub-layer in the proposed U-Transformer. In four sub-layers, $d_{\text {layer}}=512, 256, 128, 64$, respectively.}\centering
\end{figure}

The proposed attention block has three frequency-band aware multi-head attention mechanisms. The input of the block is a $T\times F\times d_{\text {layer}}$ attention map which is divided into two sub attention maps by a 1-D convolution to shuffle the features in time- and frequency-axis. The frequency attention map is further divided into two $ \frac{F}{2}\times d_{\text {layer}}$ sub-maps based on the critical frequency as 4000 Hz. The band whose frequency [4000 - 8000 Hz] is assumed as the high frequency-band and trained with fewer computation costs because it includes limited speech signals. The other frequency-band, i.e., the lower band occupies [0 - 4000 Hz] which consists of almost all the speech signals in the mixture and is the major focus of the frequency-band aware attention.

In the proposed T-F attentions method, we use eight-head attentions for both time and frequency directions. However, in the frequency-band aware attention method, the numbers of the heads in TA, HFA, and LFA are set to 8, 2, and 16 heads, respectively. In the conventional multi-head attention \cite{atyn,ATMT}, the number is set to 8 for the full frequency-band of the spectrogram. Because the desired speech signal intensively distributions at the lower frequency-band [0-4000 Hz] where requires more computation costs, LFA is trained with 16-head attention and independent learnable vectors as time attention, however, in HFA, we only exploit an overall vector $r_{c-p}^{hf}$ and the output of the multi-head attention is presented as:
\begin{equation}
W_{hf}^{Q} = \sum_{c \in \mathcal{N}_{1 \times x}(c)} (Q_{p}^{hf} K_{p}^{hf}+K_{c}^{hf})r_{c-p}^{hf}
\end{equation}
\begin{equation}
W_{hf}^{K} = \sum_{c \in \mathcal{N}_{1 \times x}(c)} Q_{p}^{hf} r_{c-p}^{hf}
\end{equation}
\begin{equation}
W_{hf}^{V} =\sum_{c \in \mathcal{N}_{1 \times x}(c)} (Q_{p}^{hf} K_{p}^{hf}+Q_{p}^{hf} r_{c-p}^{hf}+K_{c}^{hf} r_{c-p}^{hf})V_{c}
\end{equation}
Three multi-head attention blocks are trained with different sub attention maps of the extracted features and provide a combined and masked attention map which shares the same size as the feature map. The integrated attention map is added to the layer normalization with a residual connection. The pseudo-code of the high frequency-band attention is summarized as Algorithm 2.

\begin{algorithm}

  
  \SetKwInOut{Input}{input}\SetKwInOut{Output}{output}

  \Input{High frequency-band attention map as $ \frac{F}{2}\times d_{\text {layer}}$, learning rate $\eta$, epoch $E_{\max }$, Estimated LFA $M_{lf}$, time attention map $M_{t}$ from Algorithm 1 }
  \Output{Attention map as $T\times F\times d_{\text {layer}}$}
  \BlankLine
  Initialize learning vectors\;
  \For{$E = 1, 2, ...,$  $E_{\max }$}{
    \For{each column $f \in[1, \frac{F}{2}]$}{
      Train $r_{c-p}^{hf}$\;
      $W_{hf}^{O}$, $W_{hf}^{Q}$, $W_{hf}^{K}$, $W_{hf}^{V}\leftarrow r_{c-p}^{hf}$for Eq. (3)\;
      Update the frequency attention map $M_{hf}$\;
    }
    $M = M_{hf}+M_{lf}+M_{t}$ \quad //integrate three sub attention maps \; 
     
   }
  \caption{High Frequency-band Attention Algorithm.}\label{algo_disjdecomp}
\end{algorithm}
\section{4\quad Experimental Results}
In practical scenarios, the interferences consist of both background noises and undesired speech signals. Therefore, we evaluate the proposed methods in speech enhancement task with different interferences including speech and noise. The results of comparison and ablation experiments show the effectiveness of the proposed methods.
\subsection{Datasets}
We extensively perform experiments on four public datasets, including Demand \cite{Demand}, IEEE \cite{IEEE}, TIMIT \cite{TIMITs} and VOICE BANK (VCBK) \cite{VB} datasets.
\subsubsection{\quad Demand} Diverse Environments Multichannel Acoustic Noise Database (Demand) provides a set of recordings from real-world noise. We randomly collect and use 6 of 15 recordings in speech enhancement experiments with noise interferences, and the noises are $psquare$, $dliving$, $dkitchen$, $nriver$, $tcar$ and $pstation$. Each noise interference has a unique example and lasts four minutes long, and it is divided into two clips with an equal length. One is used to match the lengths of the speech signals to generate training data and another is used to generate development and inference data.
\subsubsection{\quad IEEE $\&$ TIMIT} The IEEE dataset contains speech data of American English speakers. The TIMIT dataset contains broadband recordings from 630 speakers of eight major dialects of American English, each reading ten phonetically rich sentences. In the speech enhancement experiments with speech interferences, 600 recordings from 60 speakers are randomly selected as interferences in each dataset.
\subsubsection{\quad VCBK} The VOICE BANK dataset already constitutes the largest corpora of British English. We select 1200 speech signals from 200 speakers and generate clips with an equal length as noise interferences. The generated 1200 utterances of noisy speech are then divided into 1000, 100, and 100 utterances for the training, development, and inference stages, respectively. Each utterance is divided into two two-minute clips and used in different stages.
\subsection{Baselines, Model Configuration and Performance Measurements}
In this work, three baseline models, including U-net \cite{unet}, ResU-net \cite{naagn}, and SETransformer \cite{setf} adding attention \cite{attention0} are implemented for comparison and ablation experiments.

Both U-net and ResU-net baselines use 1D convolution and zero-padded blocks. Different from the U-net, the downsampling and upsampling blocks are constructed as residual units in ResU-net. The ResU-net consists of an identity mapping and two 1-D convolution blocks. Each convolution block includes a dilated convolution layer, a batch normalization (BN), and a Leaky ReLU activation function. And the dilation of dilated convolution is applied to both the time direction and the frequency direction, which can aggregate contextual information over both time and frequency dimensions. The identity mapping with 1x1 convolution connects the input and output of the unit, which is only used to ensure the same dimensions of two tensors that pass to an addition operation \cite{naagn}.

Moreover, both the baselines and proposed methods are trained by using the Adam optimizer with a learning rate of 0.0008. All the models are initialized with normalized initialization, and the total number of epochs is 100. The dimensions of input and output magnitude spectrogram. All the experiments are run on a work station with four Nvidia GTX 1080 GPUs and 16 GB of RAM.

To evaluate and compare the quality of the enhanced speech with various methods, we use three intelligibility metrics, the short-time objective intelligibility (STOI), perceptual evaluation of speech quality (PESQ), and frequency-weighted segmental signal-to-noise ratio (fwSNRseg). The STOI and the PESQ are bounded in the range of [0, 1] and [-0.5, 4.5], respectively \cite{PESQ}. The fwSNRseg is estimated by computing the segmental signal-to-noise ratios (SNRs) in each spectral band and summing the weighed SNRs from all bands \cite{fw} in the range of [-10, 35] dB. Higher values of these measurements imply that the desired speech signal is better enhanced.

\subsection{Evaluations with Noise Interferences}
In these experiments, the averaged STOI, PESQ, and fwSNRseg performances compared with baselines using the VCBK and the Demand corpora with three SNR levels (-5, 0, 5 dB) and six noise interferences are presented in Table 1. The background noise is used to generate the speech mixture. The proposed T-F attention and frequency-band aware attention are abbreviated as TF and FAT, respectively.
\begin{table}[htbp!]
\centering
\small\addtolength{\tabcolsep}{-1pt}
\caption{Speech enhancement performance with noise interferences comparison with baselines on the \textbf{Demand} Dataset.}
\begin{tabular}{ccccccc}
\hline
Method & STOI ($\%$)  & PESQ  & fwSNRseg (dB)\\    
 \hline
 Unprocessed  & 45.6&1.58 &3.35 & \\
 DNN & 80.9&2.30 &14.35 & \\
U-net  &71.3 & 1.96 & 10.66  \\
Unet+Attention  &71.9 & 2.01 & 11.14 \\
ResU-net  &74.1 &2.04  & 12.18 \\
ResU-net+Attention &75.0 & 2.09 & 12.87\\
Transformer&78.7 & 2.19 & 13.07 \\
Transformer+Attention  &78.9 & 2.19 & 14.23 \\
\hline
\textit{Unet+TF}   &74.0 & 2.13 & 13.63 \\ 
\textit{ResU-net+TF}  &81.2 &  2.23 &  14.83\\
\textit{Transformer+TF}  &81.1 &  2.28 &  14.30\\
\textit{U-Transformer+TF}  &82.4 & 2.35 &  14.98\\
 \hline 
 \textit{Unet+FAT}  & 82.9 & 2.39 &  14.47\\
  \textit{ResU-net+FAT} & 84.3 & 2.45 &  15.53\\
 \textit{Transformer+FAT}  &84.5 &  2.49 &  16.02\\ 
  \textit{U-Transformer+FAT}  &{\bfseries 87.4} &{\bfseries 2.69} & {\bfseries 17.62}\\
 \hline 
\end{tabular}
\end{table}

Table 1 provides the averaged evaluation results. From Table 1, it can be observed that: (1) The conventional attention block has limited improvement for speech enhancement performance. However, the proposed T-F attention and frequency-band aware attention significantly improve the inference performance in all standard models. In terms of PESQ, the proposed T-F attention and frequency-band aware attention obtain 7.3$\%$ and 22.8$\%$ improvements compared with the standard Transformer model, respectively. The proposed T-F attention mechanism adopts both global connection and efficient computation on time and frequency directions. Besides, the learnable vectors $r_{c-p}^{V}$, $r_{c-p}^{K}$, and $r_{c-p}^{Q}$ for $V$, $K$, and $Q$ utilize the positional information between the positions $(p, c)$ to updates the weights $W^{V}$, $ W^{K}$, and $ W^{Q}$, respectively. The 2-D attention map is split into two 1-D sub-maps in time- and frequency-axis, which allows the parallel calculation to facilitate training \cite{ATMT}. (2) In all evaluated models, the proposed frequency-band aware attention U-Transformer shows the best effectiveness. The reason is that the proposed U-Transformer inherits advantages from both ResUnet and Transformer. Moreover, the desired information at the lower frequency-band is fully used by a 16-head attention and independent learnable vectors. However, in HFA, only an overall learnable vector $r_{c-p}^{hf}$ is applied to further improve the efficiency.

\subsection{Evaluations with Speech Interferences}
To further evaluate the proposed methods in a more challenging scenario, the undesired speech signals are randomly selected from the TIMIT and the IEEE corpora to generate mixtures. In these experiments, the averaged STOI, PESQ, and fwSNRseg performances are compared with baselines with three SNR levels (-5, 0, 5 dB) and six noise interferences and presented in Tables 2 $\&$ 3.
\begin{table}[htbp!]
\centering
\small\addtolength{\tabcolsep}{-1pt}
\caption{Speech enhancement performance with speech interferences comparison with baselines on the \textbf{TIMIT} dataset.}
\begin{tabular}{ccccccc}
\hline
Method & STOI ($\%$)  & PESQ  & fwSNRseg (dB)\\    
 \hline
 Unprocessed  & 41.5&1.44 &3.04 & \\
  DNN  & 76.9&2.06 &12.17 & \\
U-net  &67.3 & 1.88 & 8.00  \\
Unet+Attention  &67.8 & 1.97 & 8.42 \\
ResU-net &72.1 &2.09  & 9.50 \\
ResU-net+Attention &73.0 & 2.08 & 9.38\\
Transformer  &77.9 & 2.25 & 12.70 \\
Transformer+Attention  &78.3 & 2.33 & 12.84 \\
\hline
\textit{Unet+TF}   &70.4 &1.99 & 9.84 \\ 
\textit{ResU-net+TF}  &78.6 &  2.31 &  13.22\\
\textit{U-Transformer+TF}  &80.7 & 2.39 &  14.05\\
 \hline 
 \textit{Unet+FAT}  & 75.2 & 2.16 &  10.62\\
  \textit{ResU-net+FAT} & 80.7 & 2.40 &  13.16\\
  \textit{U-Transformer+FAT}  &{\bfseries 82.6} &{\bfseries 2.59} & {\bfseries 14.81}\\
 \hline 
\end{tabular}
\end{table}

Table 2 compares speech enhancement performance with the TIMIT dataset. The proposed U-Transformer model and frequency-band aware attention show significant improvements over baselines. In terms of STOI, our proposed T-F attention and frequency-band aware attention earn 2.8$\%$ and 4.7$\%$ improvements compared with the standard Transformer model, respectively. On the other hand, the experimental results suffer a degradation compared with speech enhancement performance in Table 1. It is highlighted that, due to different distributions between speech and noise interference domains, the task of personalized speech enhancement from the mixture with undesired speech signals is more challenging than from noise interferences \cite{dav}.

\begin{table}[htbp!]
\centering
\small\addtolength{\tabcolsep}{-1pt}
\caption{Speech enhancement performance with speech interferences comparison with baselines on the \textbf{IEEE} dataset.}
\begin{tabular}{ccccccc}
\hline
Method & STOI ($\%$)  & PESQ  & fwSNRseg (dB)\\    
 \hline
 Unprocessed  & 42.3&1.52 &3.11 & \\
  DNN & 79.7&2.20 &13.87 & \\
U-net &68.2 & 1.93 & 8.81  \\
Unet+Attention  &68.5 & 1.92 & 8.96 \\
ResU-net  &72.6 &2.14  & 10.77 \\
ResU-net+Attention &74.0 &2.20  & 12.51\\
Transformer  &78.4 & 2.32 & 13.08 \\
Transformer+Attention  &78.8 & 2.41 & 13.24\\
\hline
\textit{Unet+TF}   &72.0 &1.99 & 11.46\\ 
\textit{ResU-net+TF}  &80.5 &  2.38 &  13.21\\
\textit{U-Transformer+TF}  &81.6 & 2.41 &  15.47\\
 \hline 
 \textit{Unet+FAT}  & 78.1 & 2.21 &  12.55\\
  \textit{ResU-net+FAT} & 82.6 & 2.49 &  14.92\\
  \textit{U-Transformer+FAT}  &{\bfseries 85.0} &{\bfseries 2.74} & {\bfseries 17.39}\\
 \hline 
\end{tabular}
\end{table}

It can be observed from Table 3 that the speech enhancement performance is further improved by the proposed methods as evaluated in the IEEE dataset. The proposed T-F attention and frequency-band aware attention obtain 2.4 dB and 4.3 dB improvements compared with the standard Transformer model in fwSNRseg, respectively. 

Furthermore, the visualizations are given in Figure 4 which are related to the estimated spectra of the desired speech signals from different methods. After comparing the estimated spectra with the spectrogram of target speech signal, it can be observed that the spectrogram obtained via the proposed U-Transformer with frequency-band aware attention is more closer to the clean speech signal, which again confirms that the frequency-band aware attention U-Transformer method outperforms the baselines.
\begin{figure}[h!]
\centering
\includegraphics[width=8cm, height=11cm]{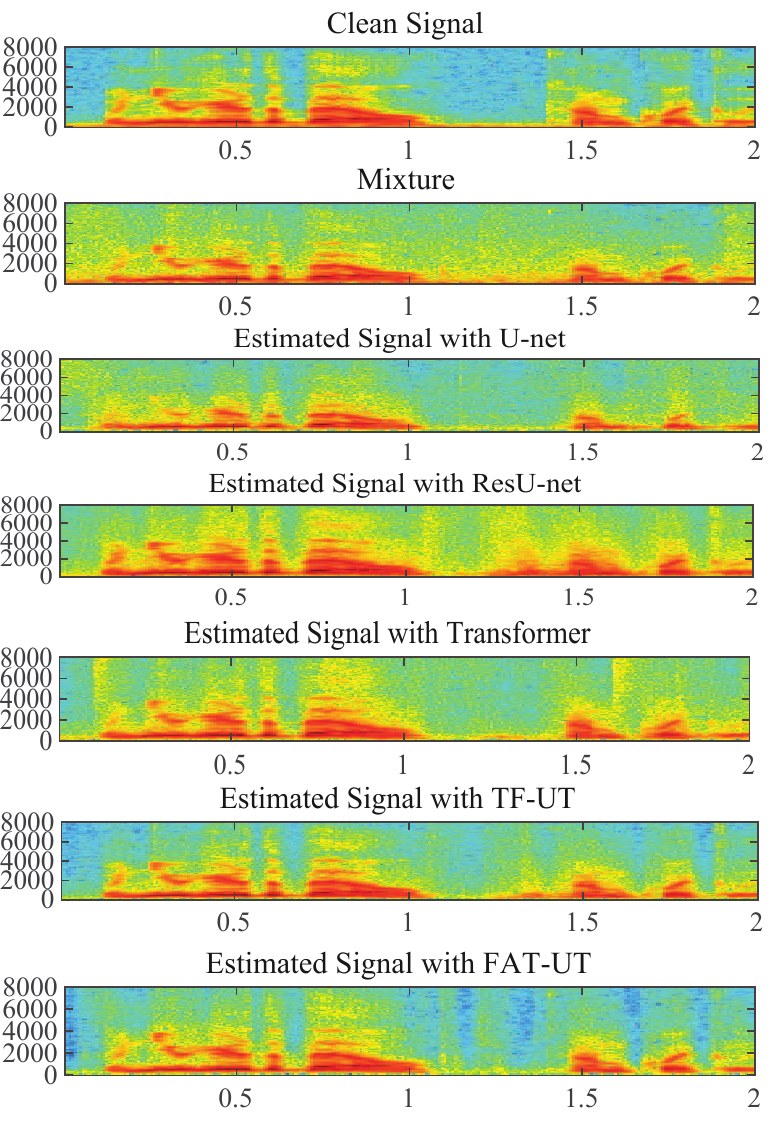}
\caption{The spectra of different signals: \textit{TF} refers to T-F attention and \textit{FAT} denotes the proposed frequency-band aware attention method. The x- and y-axis are time (s) and frequency (Hz), respectively. The inference experiment is implemented with $dliving$ and -5 dB SNR level.}\centering
\end{figure}


The above experimental results confirm that the proposed U-Transformer with the frequency-band aware attention can further improve speech enhancement performance both with noise and speech interferences compared with the baselines. In all comparison and ablation experiments, it can be observed that: (1) The U-Transformer method demonstrates that very good improvements are realized in the comparison experiments. (2) Both T-F and frequency-band aware attentions significantly outperform the conventional attention mechanism. (3) According to the ablation experiments in Tables 1-3, the frequency-band aware attention could further improve the speech enhancement performance compared with T-F attention. (4) The proposed U-Transformer with the frequency-band aware attention achieves the best enhancement performance with other state-of-the-art comparisons.  

\section{5\quad Conclusion}
In this work, we proposed a novel U-Transformer with the frequency-band aware attention for speech enhancement problems. The T-F attention split the feature map obtained from the previous sub-layer to the time and frequency directions and exploited the multi-head attention to mask sub attention maps. Consequently, the 2-D attention map was factorized into two 1D attentions and allowed parallel computations. Moreover, in order to fully use the information of the desired speech signal, the frequency-band aware attention was proposed to further split the full band into two sub-bands and different learning vectors were allocated to TA, HFA, and LFA, respectively. The experimental results confirmed that the proposed U-Transformer outperformed the state-of-the-art models and the frequency-band aware attention could help to achieve further performance improvement.

\bibliography{Formatting-Instructions-LaTeX-2022}

\end{document}